# Circuit Diagram Retrieval Based on Hierarchical Circuit Graph Representation

Ming Gao, Ruichen Qiu, Zeng Hui Chang, Kanjian Zhang, Haikun Wei and Hong Cai Chen, *Member, IEEE*

*Abstract*—In the domain of analog circuit design, the retrieval of circuit diagrams has drawn a great interest, primarily due to its vital role in the consultation of legacy designs and the detection of design plagiarism. Existing image retrieval techniques are adept at handling natural images, which converts images into feature vectors and retrieval similar images according to the closeness of these vectors. Nonetheless, these approaches exhibit limitations when applied to the more specialized and intricate domain of circuit diagrams. This paper presents a novel approach to circuit diagram retrieval by employing a graph representation of circuit diagrams, effectively reformulating the retrieval task as a graph retrieval problem. The proposed methodology consists of two principal components: a circuit diagram recognition algorithm designed to extract the circuit components and topological structure of the circuit using proposed GAM-YOLO model and a 2-step connected domain filtering algorithm, and a hierarchical retrieval strategy based on graph similarity and different graph representation methods for analog circuit. Our methodology pioneers the utilization of graph representation in the retrieval of circuit diagrams, incorporating topological features that are commonly overlooked by standard image retrieval methods. The results of our experiments substantiate the efficacy of our approach in retrieving circuit diagrams across of different types.

## I. INTRODUCTION

In the analog circuit design [1], circuit diagrams serve as the fundamental material, offering engineers a visual representation that facilitates comprehension of the circuit's operational mechanics. During the design phase, engineers often seek to consult prior designs [2, 3], the majority of which are archived as circuit diagrams. This necessitates a manual review and reproduction of these diagrams, which is not only laborious but also prone to inefficiency and error. Moreover, when engineers endeavor to identify analogous circuits for guidance in their design process, they must ensure that their creations do not infringe upon existing designs, thus avoiding potential legal complications. This requires a carefully manual examination of the designs, which is a time-consuming task. The manual retrieval of specific circuit diagrams from a vast repository is, therefore, a process that requires a significant cost of time and resources. Given these challenges, there is a demand for the development of circuit diagram retrieval algorithms. Such algorithms should be capable of recognizing similarities between designs through circuit diagrams, thereby accelerating the search process and reducing the load on engineers.

To our knowledge, few studies focus on circuit diagram retrieval have been reported. Current works mainly work on circuit component identification [4-7]. Given that circuit diagrams are in the form of image, it is logical to refer to image retrieval methods [8]. Image retrieval has been extensively explored, involving the process of retrieving images that closely resemble a query image from a dataset [9]. Both feature engineering and deep learning techniques have been developed for better image similarity comparison. However, these methods have demonstrated limited applicability when extended to the retrieval of more specialized and intricate diagram images such as circuit diagrams [10]. One major reason is that circuit diagrams exhibit a variety of forms and a single type of circuit can be represented through multiple methods of depiction. When circuit diagrams are processed solely as images without physical meaning, the topological information in the circuit diagrams may be inadvertently neglected, which is of great significance in determining the function of circuit since similar components can form circuits of completely different functions based on their distinct topological structures.

Recognizing the inherent graph structure of circuit [11-13], numerous studies have employed graph data structure to represent circuit for Electrical Design Automation (EDA). For instance, a system symmetry detection method based on graph similarity for electronic circuits has been proposed [14]; a graph-based methodology for identifying of subcircuits is introduced in [15]; graphs are used to predict circuits' output performance [16]. Furthermore, a knowledge-based strategy for generating new circuits by integrating existing circuit designs, represented as graphs, was introduced in [17]. Graph-

This research work is supported by Fundamental Research Funds for the Central Universities (No.3208002309A2) and by the Big Data Computing Center of Southeast University. *(Corresponding author: Hong Cai Chen).*

Ming Gao, Ruichen Qiu, Zeng Hui Chang, Kanjian Zhang, Haikun Wei and Hong Cai Chen are with School of Automation, and Key Laboratory of Measurement and Control of Complex Systems of Engineering, Ministry of Education, Southeast University, Nanjing, 210096, China. (e-mail: 220221874@seu.edu.cn, 220211892@seu.edu.cn, 213210618@seu.edu.cn, kjzhang@seu.edu.cn, hkwei@seu.edu.cn, chenhc@seu.edu.cn).

based deep learning models have also shown significant promise in EDA applications [18]. For instance, Graph Neural Networks (GNNs) have been effectively utilized to predict net parasitic and device parameters, as detailed in [19]. Moreover, GNNs is employed for the reverse engineering of gate-level netlists has been presented in [20]. A GNN-based approach for automated circuit topology generation and device sizing was proposed in [13]. In [21], integrated circuit designs are transformed into graphs and a GNN is used to analyze their structural similarities for detecting Intellectual Property (IP) infringement. The graph representation of circuit diagrams offers the distinct advantage of explicitly reflecting the interconnections among all elements and preserving the critical topological features and it also enables the application of graph retrieval algorithms.

This work intends to retrieve circuit diagrams that are similar to the query one from the database, in order to accelerate the process of consulting legacy designs and plagiarism check for circuit designers. The main contributions of this paper are summarized as follows:

- We introduce a new methodology for circuit diagram retrieval by representing circuit diagrams as graphs and thus transforming the retrieval process into a graph retrieval task.
- We provide a new approach to extract circuit components and circuit topology relationship from circuit diagram and generate corresponding graph representation.
- In order to achieve a balance between retrieval accuracy and retrieval time, we present a hierarchical circuit diagram retrieval method based on graph similarity and different graph representations employed in EDA.

The rest of the paper is organized as follows: In Section II, the definition and classification of circuit graph representation are introduced. In Section III, the methodology of circuit diagram retrieval based on circuit graph representation is presented. In Section IV, experimental validation of the proposed methodology is given, followed by the conclusion part in Section V.

## II. Circuit Graph Representations

### A. Definition

Graph is a data structure for representing interactions between related objects. Mathematically, it is as a tuple $G= (V, E)$, where $G$ is the set of nodes and $E$, the set of edges. The edges are defined as the connections between nodes. A graph may also incorporate labels for both nodes and edges, providing a means to encode additional information. Graphs can be used to model a wide variety of real-world problems of connections or relations, including social networks, transportation networks, communication networks and also circuits.

In the context of circuit, a natural graph representation emerges by considering electronic components as nodes and the interconnections, or nets, between these components as edges. The process of creating a graph representation of a circuit involves a systematic transformation of the circuit's constituent elements into graph's constituent elements. This includes transforming devices (e.g., transistors, passive components, hierarchical blocks), device pins (e.g., the drain, gate, and source terminals of a transistor), nets (wires), connections, device types, and pin types into corresponding nodes, edges, node labels, and edge labels within the graph representation. Fig. 1 illustrates this transformation. The construction process of the graph representation is as follows: represent 2 MOS transistors M1, M2 in the circuit as nodes M1, M2 in the graph, and represent 3 nets as nodes D1, D2, S; add edges between the nodes according to the connections between the MOS transistors and nets.

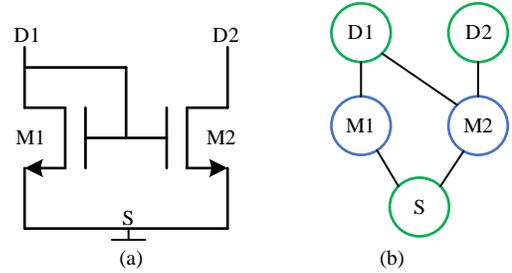

Fig. 1. Transformation from circuit to graph representation. (a) Circuit diagram. (b) Graph representation.

### B. Classification of Circuit Graph Representations

In the domain of EDA, several researches have been conducted premised on the conceptualization of circuits as graphs. These studies have utilized a range of graph representations to model and analyze circuits. The complexity and informational fidelity of a graph representation are closely connected to the number of nodes and edges that the graph contains. It is logical to infer that a graph with fewer nodes and edges tends to be simpler but captures fewer details of the original circuit.

The number of nodes and edges in a graph representation is significantly influenced by the graph construction method employed. This method is defined as the correspondence between elements of the circuit and the nodes and edges within the graph representation, which varies among the reviewed graph representations. Consequently, the classification of these graph representations is based on their distinct graph construction methods. The reviewed graph representations can be systematically categorized into 5 classes, which is shown in Table I.

TABLE I
CLASSIFICATION OF GRAPH REPRESENTATIONS REVIEWED

| Class | Ref. | Node Construction | Edge Construction |
|---|---|---|---|
| Class 1 | [17, 22-27] | Device | Net |
| Class 2 | [28] | Device, net (connected with IO pin) | Net (not connected with IO pin) |
| Class 3 | [15, 19, 29-37] | Device, net | Connection |
| Class 4 | [16, 38-40] | Device, device pin | Connection |
| Class 5 | [41, 42] | Net | Device |

Class 1: The most straightforward graph construction method is to represent the devices in the circuit as nodes in the graph, and represent the nets in the circuit as edges in the graph, which is adopted in [17, 22-27]. In this class, each node is attached with a node label decided by the corresponding device type. In order to identify the pin connections of each edge, the connections of different pins between two devices are given different scores, and the edge label of the edge between two devices equals to the sum of all the scores of the connected pins between the two devices. Fig. 2 (b) shows the transformation from circuit to graph representation. The construction process of the graph representation is as follows: represent 2 MOS transistors in the circuit as nodes M1, M2 in the graph; add edges between the nodes according to the connections between the MOS transistors and nets; calculate the edge label of each edge according to pin connections. For simplicity, the labels are not shown.

Class 2: In [28], a circuit is encoded into a graph, in which devices and nets (connected with IO pins) are represented as nodes and nets (not connected with IO pins) are represented as edges. The transformation from circuit to graph representation is illustrated in Fig. 2 (c). The construction process of the graph representation is as follows: represent 2 MOS transistors and their shared source pin as nodes M1, M2, S in the graph; add edges between the nodes according to the connections.

Class 3: In order to represent circuit more explicitly, both devices and nets are represented as nodes and connections between devices and nets are represented as edges in [15, 19, 29-37].In this class, there are no edges between the nodes of devices and nodes of circuit nets, so the graph created is bipartite. Moreover, edges of the graph are assigned with three-bit labels determined by the type of device pins connected to circuit nets as edge feature. Fig. 2 (d) shows the transformation from circuit to graph representation. The construction process of the graph representation is as follows: represent 2 MOS transistors in the analog circuit as nodes M1, M2 in the graph, represent 3 nets as nodes D1, D2, S; add edges between the nodes according to the connections between the MOS transistors and nets; add the edge label of each edge according to pin connections.

Class 4: In [16, 38-40], both devices and device pins are represented as nodes and nets are represented as edges to include information of pin connections directly in the topological structure of the graph representation. In this class, edges connect device nodes to their corresponding device pin nodes and device pin nodes are connected in the graph accordingly if there exist nets between them. The transformation from circuit to graph representation is illustrated in Fig. 2 (e). The construction process of the graph representation is as follows: represent 2 MOS transistors as nodes M1, M2 and represent all the transistor pins in the circuit as nodes G1, G2, S1, S2, D1, D2 in the graph; add edges between the nodes according to the connections between the transistors and transistor pins.

Class 5: In [41, 42] , a graph is constructed to represent circuit, with its nodes being the circuit nets, and its edges being the devices. The edges link two nodes if the associated nodes are the source and drain of the same MOS transistor, respectively. Note that this class of graph representation can only be applied for CMOS and bipolar circuits. The transformation from circuit to graph representation is illustrated in Fig. 2 (f). The construction process of the graph representation is as follows: represent 3 nets as nodes D1, D2, S in the graph; add edges M1, M2 to represent 2 MOS transistors.

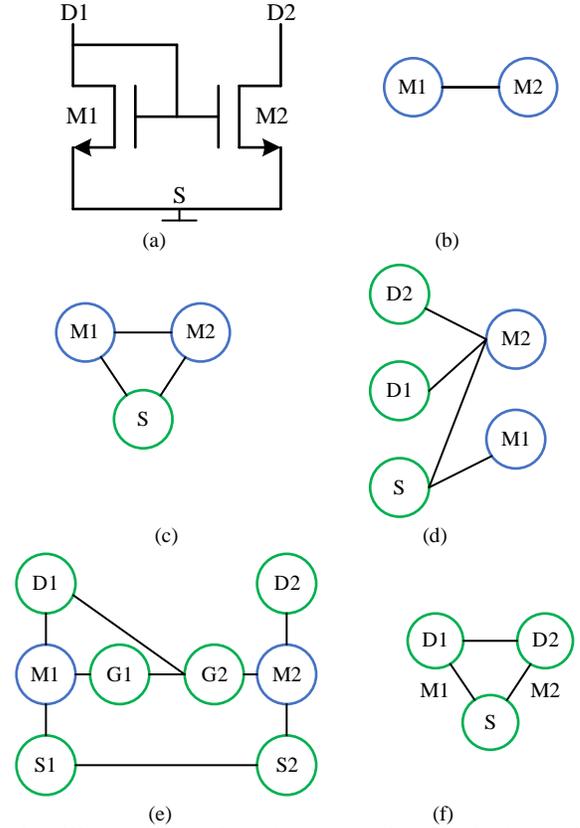

Fig. 2. Different classes of graph representation for circuit. (a) Circuit Diagram. (b) Graph representation in Class 1. (c) Graph representation in Class 2. (d) Graph representation in Class 3. (e) Graph representation in Class 4. (f) Graph representation in Class 5.

## III. METHODOLOGY

### A. Circuit Diagram Retrieval: Overview

In this paper, we propose to transform circuit diagram into graph and redefining the circuit diagram retrieval task as a graph retrieval problem. The graph representation of circuit diagrams offers clear advantages, namely it clearly reflects the mutual connections between all circuit components and retains the topological features that are very critical to determining the function of the circuit.

The whole process is divided into 2 steps, which is illustrated in Fig. 3: circuit diagram recognition and hierarchical graph retrieval. Firstly, the circuit components and circuit topology will be extracted from the circuit diagram with object detection method and edge detection algorithm respectively and after that the circuit diagram can be transformed into graph through intersection detection. Secondly, the GED (graph edit distance) is calculated as the graph similarity metric between the query graph and database graphs in order to get the most similar graphs to the query graph in the database. In order to achieve a balance between retrieval accuracy and efficiency, a hierarchical retrieval method is proposed which first roughly retrieve within the database using simpler representation (Class 1) and further retrieve within the result of the first stage using more detailed representation (Class 4).

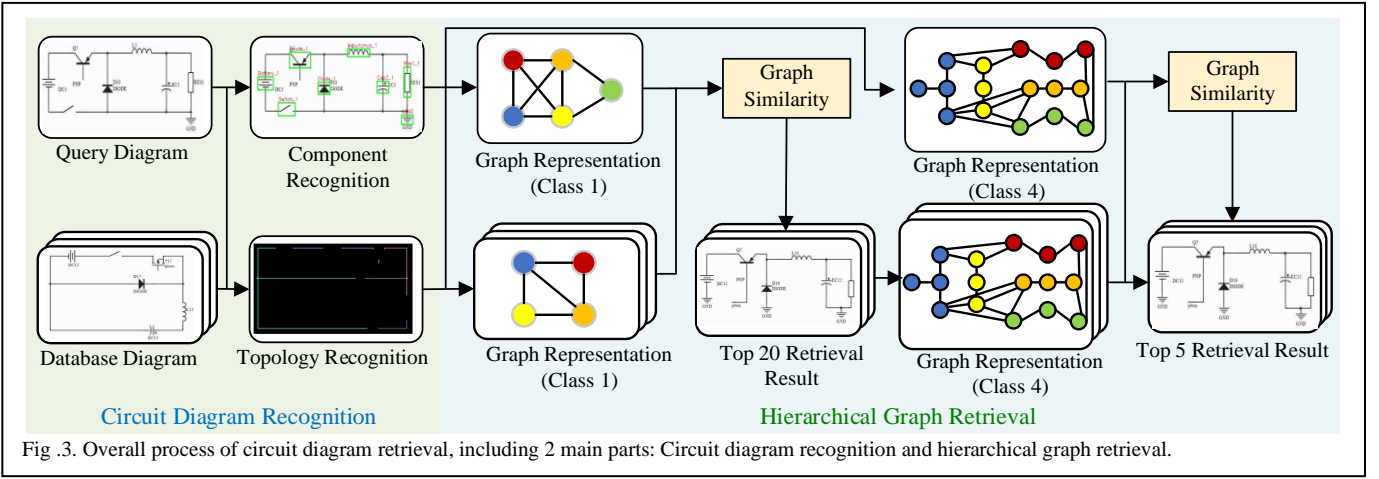

Fig .3. Overall process of circuit diagram retrieval, including 2 main parts: Circuit diagram recognition and hierarchical graph retrieval.

### B. Circuit Diagram Recognition

#### 1) Circuit Component Recognition

For circuit component recognition, an Attentional YOLOv8 model, GAM-YOLO is employed. YOLOv8 [43] is the SOTA (state of the art) model for object detection. The overall architecture of YOLOv8, can primarily be divided into 3 main parts: the backbone, the neck, and the head. Backbone is the first major component, responsible for extracting features from the original input image. The backbone consists of the Convolutional (Conv) module, the C2f module, the Spatial Pyramid Pooling Feature (SPPF) module. As for neck, the structure adopted is Feature Pyramid Network (FPN) combined with Path Aggregation Network (PAN) module for further process and fusion of the features extracted by backbone. Lastly, the head is responsible for predicting the position and category of the target based on the feature maps processed at the neck. Based on the concept of Anchor-Free, YOLOv8 decouples the detection head into a classification part and a localization part. For classification part, YOLOv8 employs VariFocal Loss (VFL) [44] as classification loss function, the equation is as follows:

$$VFL(p,q) = \begin{cases} -q(q\log(p)+(1-q)\log(1-p)) & \text{if } q > 0 \\ -\alpha p^{\gamma}\log(1-p) & \text{if } q = 0 \end{cases} \quad (1)$$

In (1), $p \in [0, 1]$ represents the predicted probability, while $q \in [0, 1]$ represents the class label related to the Intersection over Union (IoU), and $\alpha$ and $\gamma$ are hyperparameters. For localization, The YOLOv8 incorporates the Distribution Focal Loss (DFL) [45] on the basis of the Complete Intersection of Union (CIoU) localization loss function from YOLOv5, the equation is as follows:

$$DFL(p,q) = -((y_{i+1}-y)\log(S_i)+(y-y_i)\log(S_{i+1})) \quad (2)$$

In (2), $y_i$ represents the label value, $y_i$ and $y_{i+1}$ respectively represent the rounded values of $y$ on the left and right sides. $S_i$ and $S_{i+1}$ respectively represent the predicted values of the bounding box at the positions $y_i$ and $y_{i+1}$.

In the task of component detection in circuit diagrams, components often have distinct foreground and spatial features. Introducing channel attention mechanisms and spatial attention mechanisms can enable the algorithm to better recognize components. The Global Attention Module (GAM) attention mechanism is an algorithm that combines channel attention and spatial attention. In terms of overall architecture, GAM connects the channel attention module in series with the spatial attention module. The input feature map is processed sequentially by the channel attention matrix and the spatial attention matrix, the equation is as follows:

$$F3 = F1 \otimes M_c \otimes M_s \quad (3)$$

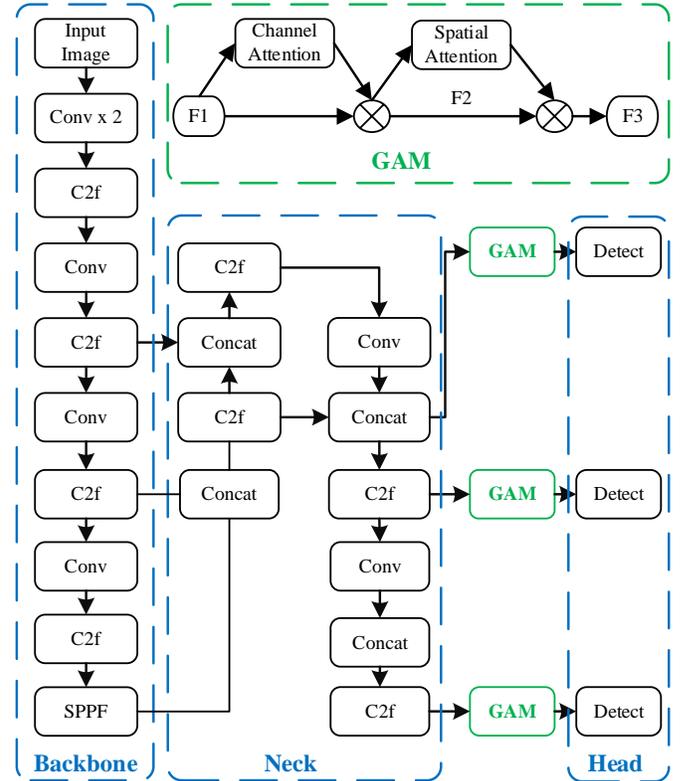

Fig.4. Overall architecture of GAM-YOLO.

In (3), *F1* and *F3* represent the input feature map and the output feature map respectively. $M_c$ and $M_s$ denote the channel attention weight matrix and the spatial attention weight matrix respectively. This paper introduces an attentional YOLOv8 model: GAM-YOLO by adding GAM mechanism between the neck and the head of the YOLOv8 to form an improved network, as shown in Fig.4, after the neck of the network outputs the three fused feature maps P1, P2, and P3, a GAM is connected in series. With this improvement, the network weights or adjusts the feature maps before generating bounding boxes and class predictions, thereby enhancing the feature representation of the regions of interest, which in turn further improves detection performance and accuracy. This method can enhance the network's ability to abstract targets and improve the detection effect on complex scenes and small targets.

### 2) Circuit Topology Recognition

Image preprocessing is the first step in circuit topology recognition, mainly including 2 steps: gray-scaling and binarization. The equation below is used as the grayscale conversion formula in this paper, where *GRAY* represents the pixel value after grayscale conversion, and *R*, *G*, *B* represent the pixel values of the red, green, and blue channels of the color image, respectively:

$$GRAY = 0.299 \times R + 0.587 \times G + 0.114 \times B \quad (4)$$

Due to the complexity of the background in circuit images, manually setting a binarization threshold cannot meet the needs of all images, and an adaptive threshold binarization algorithm is required. The most common adaptive threshold binarization algorithms are two types: Triangle method binarization and Otsu's method binarization. Triangle method binarization involves drawing a straight line with the peaks and valleys of the grayscale histogram as endpoints, and then finding the point on the grayscale histogram that is farthest from this line as the threshold. Otsu's method binarization is based on the core idea of maximizing inter-class variance and minimizing intra-class variance. The equation below represents the inter-class variance of Otsu's method binarization, and the grayscale value that maximizes this formula is used as the threshold for binarization:

$$\sigma^2 = P_1 P_2 (m_1 - m_2)^2 \quad (5)$$

In the equation, *P1*, *P2* represent the probabilities of pixels being divided into two classes, and *m1, m2* represent the mean values of the pixels for each of the two classes, respectively.

For circuit diagrams in light background, Otsu's method is chosen for binarization, which can directly filter out scattered points and grid backgrounds, whereas the triangle method binarization tends to introduce more noise. For circuit diagrams with a dark background, employing the triangle method for binarization can retain more information, ensuring that the dark wire information is not lost, and any noise generated, can be resolved through post-processing.

After image preprocessing, a 2-step connected domain filtering algorithm is applied to extract net information for determining circuit topology. The first connectivity domain filtering removes connectivity domain that are smaller than 10% of the image's pixel points, thereby filtering out a large amount of text and noise information to obtain the main body of the circuit. The second connectivity domain filtering requires the addition of component location information, and if the detected connectivity domain does not intersect with any predicted component bounding box, the connectivity domain is filtered out. The remaining connectivity domains are equivalent to the nets of the circuit. An example of the process is illustrated in Fig.5. After getting circuit component information and circuit net information of the circuit from the circuit diagram, through intersection detection, the topology relationship can be generated.

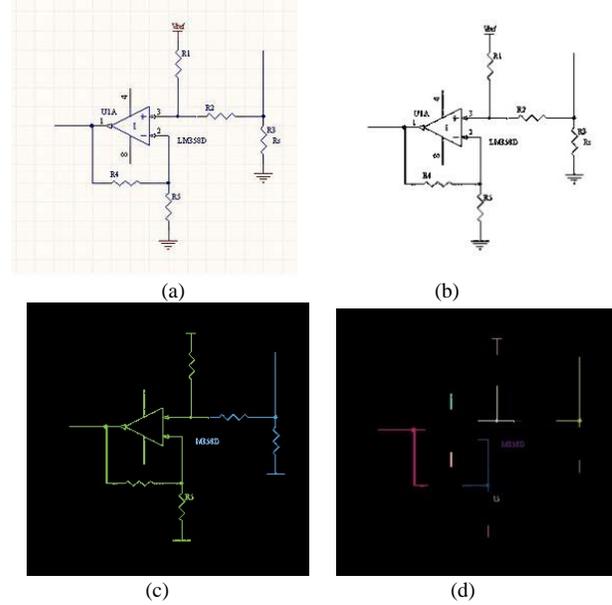

Fig.5. An example of circuit topology recognition. (a) Original circuit diagram. (b) After preprocessing (c) After the first filtering. (d) After the second filtering.

### C. Hierarchical Graph Retrieval

#### 1) Graph Retrieval Based on Graph Similarity

Graph retrieval involves the process of identifying the most relevant or structurally similar graphs within a database to a given query graph. A critical step in this process is the computation of similarity scores between the query graph and each graph in the database, followed by the ranking of the top *K* most similar graphs. The preservation of the original circuit diagram's information within its graph-based representation ensures that the calculated graph similarity scores are indicative of the underlying similarity between the corresponding circuit diagrams. A variety of graph similarity metrics are leveraged in the domain of graph retrieval, including but not limited to GED (graph edit distance) [46], and MCS (maximum common subgraph) [47].

In this paper, GED is chosen as the graph similarity metric. The GED between two graphs is defined as the minimum number of edit operations that can transform one graph to the other graph, where the edit operations are edge insertion /deletion/relabeling and node insertion/deletion/relabeling. GED gives the minimum amount of distortion needed to transform one graph into the other [48]. Consider the two graphs $q$ and $g$ in Fig. 6, where node labels are illustrated inside circles (i.e., A, B) and edge labels are illustrated beside edges (i.e., a, b). One possible sequence of edit operations for transforming $q$ into $g$ is as follows: (1) change the label of edge (v1, v2) from b to a, (2) delete edge (v1, v3), and (3) insert edge (v2, v4) with label b. Thus, the GED between $q$ and $g$ is at most 3.

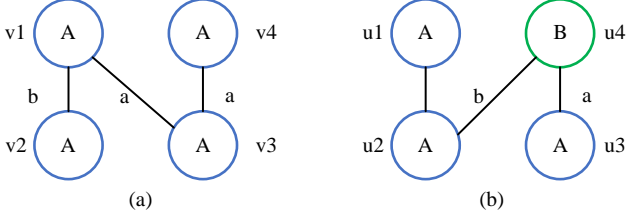

Fig .6. Sample graphs for computing GED. (a) Graph $q$, (b) Graph $g$.

In this paper, we employ enhanced $A^*$ algorithm [48] to calculate the GED value between two graphs: $GED(G1, G2)$. Since the goal is to achieve the similarity score rather than the GED value between two graphs, normalization is applied to GED . The mathematical description is provided in equations below:

$$nGED(G1, G2) = \frac{GED(G1, G2)}{(\mid G1 \mid + \mid G2 \mid) / 2} \qquad (6)$$

$$S(G1, G2) = e^{-nGED} \qquad (7)$$

where $G1$, $G2$ represent two graphs, $|G|$ represents the number of nodes in $G$, $S(G1, G2)$ is the similarity score between the two graphs.

**2) Hierarchical Graph Representation for Retrieval**

The selection of an appropriate graph representation method for the circuit is critical to the success of the circuit diagram retrieval, which is supposed to effectively retain the original circuit information while striving for simplicity. Except for graph representation in Class 5 that can only be applied for CMOS or bipolar circuits, from Class 1 to Class 4, the graph representation becomes more complex and contains more original circuit information, leading to higher retrieval accuracy and more retrieval time. As a result, there is a need for a graph representation method that can achieve a balance between retrieval accuracy and efficiency.

In this paper, an improved hierarchical graph representation of circuit for retrieval is proposed, which involves 2 stages. At first, the simplest graph representation in Class 1 is employed to represent all the circuit diagrams in the database and the query circuit diagram for graph similarity computation and acquire top 20 retrieval result from the database. After that, the most complicated graph representation in Class 4 is employed to represent the top 20 retrieved circuit diagrams in the first step and the query circuit diagram for graph similarity computation and get top 5 retrieval result from the top 20 retrieval result.

## IV. EXPERIMENT RESULTS

This part consists of 2 experiments for the proposed method: experiment of circuit diagram recognition and experiment of circuit diagram retrieval. Finally, the proposed method is compared with traditional image-based retrieval techniques to highlight its advantages.

### A. Circuit Recognition

**1) Dataset**

For circuit component recognition, the database constructed in this paper is sourced from patent circuit images, web circuit images, datasheet circuit images, and screenshots from circuit simulation software, totaling 2275 images (including 23304 circuit components), in which 2048 images for training dataset and 227 images for testing dataset. The dataset is labeled using the labeling script. The specific labeling categories can be seen in Fig.7, and the number of instances of each category is shown in Table III.

Fig .7. Labeling categories for circuit component detection.

TABLE III
Number of Instances of Each Category

| Category Label | Number of Instances | Category Label | Number of Instances |
|---|---|---|---|
| Res1 | 1904 | DCPower | 634 |
| Res2 | 4175 | ACPower | 114 |
| Cap1 | 3040 | CurPower | 660 |
| Cap2 | 529 | Crystal | 34 |
| Inductance | 2043 | Switch | 295 |
| Diode | 2450 | Lamp | 67 |
| Bidiode1 | 53 | Speaker | 27 |
| Biodioe2 | 53 | Motor | 47 |
| Triode | 666 | DeviceA | 29 |
| Mos | 1231 | DeviceV | 10 |
| Amplifier | 218 | DeviceM | 35 |
| Trans | 181 | DeviceOsc | 25 |
| Bridge | 70 | Module1 | 225 |
| Relay | 65 | Module2 | 131 |
| AGND | 987 | notCon | 900 |
| DGND | 165 | notCon2 | 1480 |
| PGND | 22 | Optocoupler | 22 |
| Battery | 689 | Total | 23304 |

On the other hand, to verify the effectiveness of the circuit topology recognition algorithm, 250 circuit diagrams from the dataset are randomly chosen for test.

### 2) Evaluation Metrics

To evaluate the performance in the task of circuit component detection in circuit diagrams, commonly used object detection evaluation metrics are introduced. We shall first introduce the IoU (Intersection of Union). If the IoU of the object is greater than the threshold, it will be considered as true positive (TP). Otherwise, it will be considered as false negative (FN). On the other hand, if the object is not in the ground truth, but the model detects it, it will be considered as false positive (FP). Based on the IoU thresholds, we can evaluate the performance of the object detection models according to their precision, recall, and F1-score, which can be obtained from the value of TP, FN and FP. Another widely adopted evaluation metric in object detection is called mean average precision (mAP). mAP is averaged from the average precision (AP) of each object type. AP is calculated by the area under the smooth line of the precision-recall curve.

Meanwhile, the evaluation of circuit topology recognition algorithm is included in the evaluation of the overall circuit diagram recognition algorithm. Since these circuits do not have a standard topology description, this paper uses professional circuit knowledge to manually judge the generated graph. If the generated graph can accurately describe the circuit topology, the circuit diagram is considered to be correctly recognized.

### 3) Experiment Result

Table IV shows the resulting precision, recall, F1-score, and mAP for the proposed GAM-YOLO model.

TABLE IV
CIRCUIT COMPONENT DETECTION RESULT

| Number of images | 227 |
|---|---|
| Instances | 2140 |
| Precision | 0.914 |
| Recall | 0.907 |
| mAP50 | 0.902 |
| F1-score | 0.910 |

The GAM-YOLO model obtained from training is imported into the overall circuit recognition algorithm. the statistical results of the experiment are shown in Table V, and examples of recognition results of circuit diagrams under different backgrounds are shown in Fig. 8. The results show that under various backgrounds, the algorithm can accurately recognize the circuit components and their topological relationships.

TABLE V
CIRCUIT DIAGRAM RECOGNITION RESULT

| Number of diagrams | 250 |
|---|---|
| Number of accurate recognitions | 223 |
| Precision | 0.892 |

### B. Circuit Diagram Retrieval

#### 1) Experiment Details

The experiment is conducted on a database consisting of 503 circuit diagrams in 10 different types sourced from patent circuit images, web circuit images, datasheet circuit images, and screenshots from circuit simulation software.

When it comes to the selection of graph representations for experiment. The selected graph representations must demonstrate universal applicability across all circuit types. Graph representations designed for specific circuit classes, such as those applicable solely to CMOS and bipolar circuits within Class 5, are unsuitable for the experiment. As a result, graph representations in Class 1, Class 2, Class 3 and Class 4 are selected for experiment.

In addition, in order to reduce the difficulty of generating graph representations from circuit diagram, we have adopted a simplification strategy for the selected graph representations. Firstly, all graph representations are modeled as undirected. Secondly, the node label is only determined by the corresponding device or net type. Lastly, edge labels are not considered in our graph representations. This decision is motivated by the fact that edge labels, which are typically determined by the types of pins connected within the circuit, are often difficult to acquire automatically from the circuit diagram.

#### 2) Evaluation Metrics

Firstly, average number of nodes ($N_n$) and average number of edges ($N_e$) in the graph representations of the circuit diagram database are calculated to show the simplicity of each graph representation. Meanwhile, the performances of different graph representations are evaluated by speed and accuracy. As for speed, the average time spent on retrieval for each query circuit diagram in the database is calculated. The mathematical description is provided in the following equation, where $AT$ represents average retrieval time of each query circuit diagram, $T$ represents total retrieval time of all the circuit diagrams in the database, $N$ represents the total number of circuit diagrams in the database:

$$AT = \frac{T}{N} \tag{8}$$

As for accuracy, the circuit diagrams are of different types and if the retrieved circuit diagram and the query circuit diagram are of the same type, the retrieval is accurate. The accuracy is evaluated by calculating the average precision of the top 5 retrieval result of all the query circuit diagrams. The mathematical description is provided in the following equation, where $P_i^K$ represents the top K retrieval precision of the query circuit diagram $i$, $N$ represents the total number of circuit diagrams in the database, and $AP_{retrieval}$ represents the average precision of the top 5 retrieval result of all the query circuit diagrams:

$$AP_{retrieval} = \frac{\sum_{i=1}^{N}\sum_{K=1}^{5}P_i^K}{N} \tag{9}$$

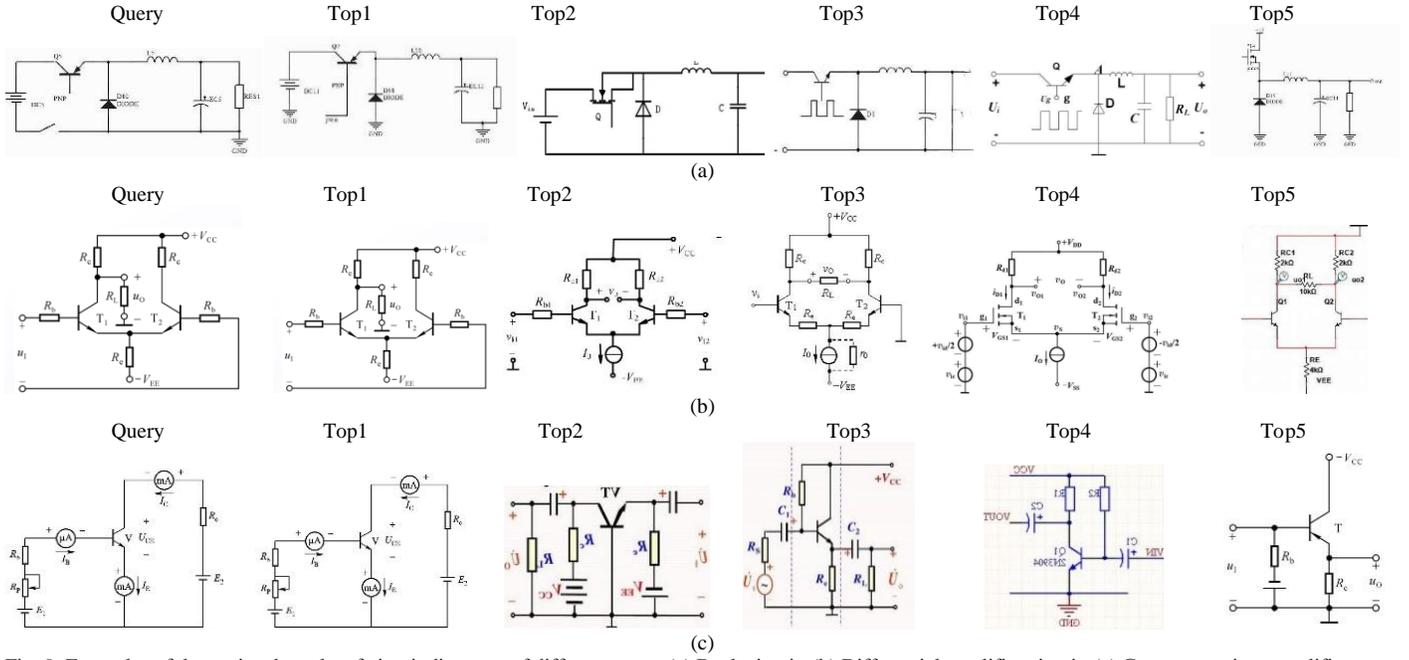

Fig. 9. Examples of the retrieval results of circuit diagrams of different types. (a) Buck circuit. (b) Differential amplifier circuit. (c) Common emitter amplifier circuit.

### 3) Experiment Result

The query circuit diagram and the circuit diagrams in the database are represented as graphs through the circuit diagram recognition process and then hierarchical graph retrieval process would retrieve circuit diagrams from the database ranked by similarity to the query circuit diagram. The result of the experiment is shown is TABLE VI.

#### TABLE VI
#### CIRCUIT DIAGRAM RETRIEVAL RESULT

| Graph Representation | $N_n$ | $N_e$ | $AP_{retrieval}$ | $AT$ (s) |
|---|---|---|---|---|
| Class 1 | 6.27 | 9.53 | 0.815 | **47.64** |
| Class 2 | 8.45 | 10.32 | 0.824 | 96.81 |
| Class 3 | 12.11 | 13.29 | 0.852 | 222.03 |
| Class 4 | 13.47 | 14.60 | 0.881 | 252.15 |
| Hierarchical | 6.27(Class1)/ 13.21(Class4) | 9.53/14.35 | **0.881** | 57.86 |

As for the simplicity of the graph representations, compared with graph representations in Class1 which only represent devices as nodes, graph representations in Class 2 and Class 3 add extra nodes of nets and extra nodes of nets connected with IO pins respectively; graph representations in Class 4 add extra nodes of device pins. As a result, from Class 1 to Class 4, graph representation becomes more complex and contains more original information in the circuit diagram.

Concerning performance for retrieval task, according to the experiment result, from Class 1 to Class 4, the average precision of circuit diagram retrieval increases as the graph representation preserves more original information in the circuit diagram; the average retrieval time increases dramatically because calculating the chose graph similarity metric (GED) is a NP-complete problem and it takes much more time to calculate GED between 2 graphs when the number of nodes increase [48]. Lastly, the hierarchical representation method achieves average precision that is very close to the most complicated graph representation in Class 4, but the average retrieval time is greatly reduced, which proves the advantage of the proposed hierarchical representation method. Examples of the retrieval results of circuit diagrams of different functions are shown in Fig. 9.

### C. Comparison to Image-Based Retrieval Methods

Traditional retrieval methods are mainly based on image similarity. To prove the advantages of the proposed method, three image-based retrieval methods are evaluated to retrieve circuit diagrams for comparison [49, 50]. The details of these methods are as follows:

① ResNet101 [49] serves as the backbone network to generate a feature map, which is then condensed into a compact vector by a global-aggregation layer. This vector is subsequently processed through a Fully Connected (FC) layer and normalized using L2 normalization. Cosine distance is used to measure the similarity between images.

② VGG16 [50] is employed as the backbone network to extract features from the input images. Following feature extraction, Singular Value Decomposition (SVD) is applied to reduce the dimensionality of the feature space. Then, Query Expansion (QE) is used for combining the retrieved top-k nearest neighbors with the original query and doing another retrieval to improve performance.

③ VGG16 [50] is used as the feature extraction backbone to capture image features. These high-dimensional features are then projected into a lower-dimensional space using Principal Component Analysis (PCA). To further refine our retrieval process, QE is also used to rerank the initial retrieval results.

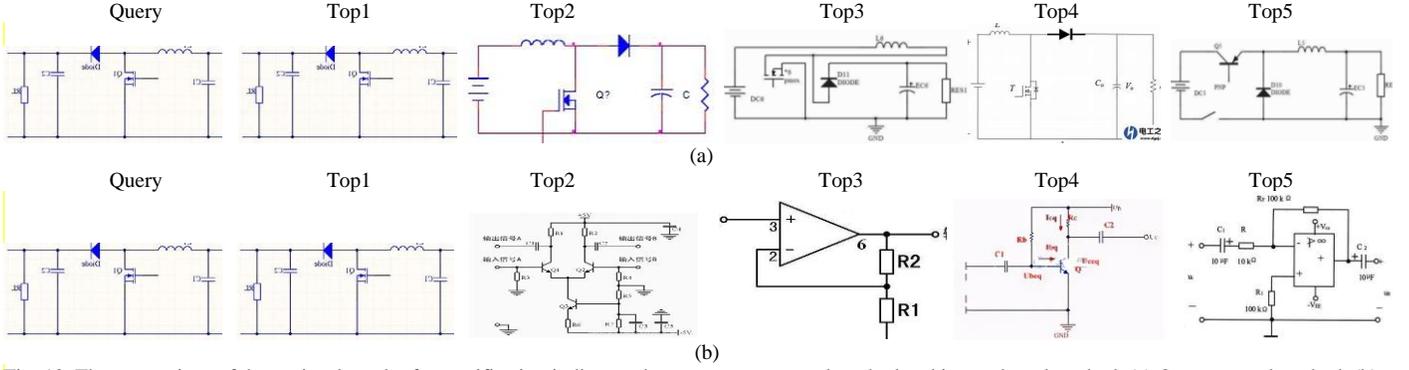

Fig. 10. The comparison of the retrieval result of a specific circuit diagram between our proposed method and image-based method. (a) Our proposed method. (b) Image-based method ①.

Table VII illustrates the comparative analysis of retrieval performance on our dataset of circuit diagrams. It contrasts our proposed graph-based retrieval method with the image-based retrieval methods referenced in [49] and [50]. The data clearly demonstrates that our graph-based approach surpasses the image-based methods in terms of retrieval accuracy.

For demonstration, the comparison of the retrieval result with image-based method ① and our proposed graph-based method of a specific circuit diagram is shown in Fig.10. It can be seen that similar images can be found using the image-based method ①. While, the same circuit with varied routing direction cannot be found even the change is small. It also proves the necessity to develop a circuit retrieval method.

TABLE VII
CIRCUIT DIAGRAM RECOGNITION RESULT

| Method | $AP_{retrieval}$ |
| --- | --- |
| image-based retrieval method ① | 0.529 |
| image-based retrieval method ② | 0.416 |
| image-based retrieval method ③ | 0.627 |
| Our graph-based method (Hierarchical) | **0.881** |

## V. CONCLUSION

In this paper, we have introduced a new methodology for circuit diagram retrieval by employing circuit graph representation and redefining the retrieval process as a graph retrieval task. We offer an innovative strategy for extracting the constituent elements and the topological interconnections of a circuit from its diagram by using proposed GAM-YOLO model and 2-step connected domain filtering algorithm, subsequently creating a corresponding graph representation. To strike a balance between the precision of the search and the time it takes to perform it, we introduce a hierarchical approach to circuit diagram retrieval that leverages graph similarity and different graph representations employed in EDA. The proposed methods are validated on a database of circuit diagrams sourced from patent circuit images, web circuit images, datasheet circuit images, and screenshots from circuit simulation software.

There are several avenues for future research that could improve the circuit diagram retrieval process. Firstly, machine learning algorithms can be applied for accelerating the calculation of graph similarity score. Secondly, considering that the impact of different components on circuit function varies, it is logical to add an algorithm to reflect the similarity between 2 graphs from node-level.